\documentclass[11pt]{article}
\usepackage{amsmath}
\usepackage{graphicx}
\usepackage{fancyheadings}
\usepackage{amsfonts}
\usepackage{amssymb}
\begin{document}

\date{}
\title{\textbf{Superfield Extended BRST Quantization in General Coordinates}}
\author{\textsc{B.~Geyer}$^{a)}$\thanks{E-mail: geyer@itp.uni-leipzig.de},
\textsc{D.M.~Gitman}$^{b)}$\thanks{E-mail: gitman@dfn.if.usp.br},
\textsc{P.M.~Lavrov}$^{a),c)}$\thanks{E-mail:\textit{\ }lavrov@tspu.edu.ru},
and \textsc{P.Yu. Moshin}$^{b),c)}$\thanks{E-mail: moshin@dfn.if.usp.br}\\ \\\textit{$^{a)}$Center of Theoretical Studies, Leipzig University,}\\\textit{Augustusplatz 10/11, D-04109 Leipzig, Germany} \\\textit{$^{b)}$Instituto de F{\'{\i}}sica, Universidade de S\~{a}o Paulo,}\\\textit{Caixa Postal 66318-CEP, 05315-970 S\~{a}o Paulo, S.P., Brazil}\\\textit{$^{c)}$Tomsk State Pedagogical University,} \textit{634041 Tomsk, Russia}}
\maketitle
\begin{abstract}
We propose a superfield formalism of Lagrangian BRST--antiBRST quantization of
arbitrary gauge theories in general coordinates with the base manifold of
fields and antifields described in terms of both bosonic and fermionic variables.
\end{abstract}

\section{Introduction}

The principle of extended BRST symmetry provides the basis of several
Lagrangian quantization schemes for general gauge theories, including the
well-known $Sp(2)$-covariant approach \cite{BLT} and its different
modifications, e.g., the superfield formalism \cite{L} and the two versions of
triplectic quantization \cite{3pl,mod3pl}. In order to reveal the geometric
content of extended BRST symmetry, it is important to study these quantization
methods in general coordinates (see, e.g., \cite{gln,gl} and references therein).

In the recent paper \cite{gl}, it was shown that the geometry of the
$Sp(2)$-covariant and triplectic schemes is the geometry of an even symplectic
supermanifold equipped with a scalar density function and a flat symmetric
connection (covariant derivative), while the geometry of the modified
triplectic quantization also includes a symmetric structure (analogous to a
metric tensor). The study of \cite{gl} generalizes the concept of triplectic
supermanifolds, introduced in \cite{gln}, to the case of base manifolds
\cite{gln,gl} containing not only bosonic but also fermionic variables.

In this paper, we propose a superfield version of the quantization scheme
developed in \cite{gln,gl}. The superfield description naturally involves an
extension of supermanifolds used in \cite{gln,gl}. Namely, the triplectic
supermanifold is extended to the complete supermanifold of variables used in
the original $Sp(2)$-covariant approach. Note that in Darboux coordinates a
similar extension takes place in the superfield formulation \cite{L} of the
$Sp(2)$-covariant scheme.

The paper is organized as follows. In Section 2, we propose a superfield
extension of triplectic supermanifolds and introduce an operation of covariant
differentiation on such supermanifolds, following the approach of our previous
works \cite{gln,gl}. In Section 3, we propose a manifest realization of the
(modified) triplectic algebra \cite{3pl,mod3pl} and outline a suitable
quantization procedure along the lines of \cite{gln,gl}. In Section 4, we
summarize the results and make concluding remarks.

We use DeWitt's condensed notation \cite{condnot} and apply tensor analysis on
supermanifolds \cite{DeWitt}. Left-hand derivatives with respect to some
variables $x^{i}$ are denoted as $\partial_{i}A={\partial A}/{\partial x^{i}}
$. Right-hand derivatives with respect to $x^{i}$ are labelled by the
subscript $"r"$, and the notation $A_{,i}={\partial_{r}A}/{\partial x^{i}}$ is
used. The covariant derivative $\nabla$ (and other operators acting on tensor
fields) is assumed to act from the right: $A\nabla$; if necessary, the action
of an operator from the right is indicated by an arrow, e.g., $\overleftarrow
{\nabla}$. Raising the $Sp(2)$-group indices is performed with the help of the
antisymmetric second rank tensor $\varepsilon^{ab}$ ($a,b=1,2$): $\theta
^{a}=\varepsilon^{ab}\theta_{b}$, $\varepsilon^{ac}\varepsilon_{cb}=\delta
_{b}^{a}$. The Grassmann parity of a quantity $A$ is denoted by $\epsilon(A)$.

\section{Superfield Extension of Triplectic Supermanifolds}

The supervariables used in various realizations of extended BRST\ symmetry can
be naturally combined into a set $(x^{i},\theta_{a}^{i},y^{i})$,
$i=1,2,\ldots$, $N=2n$. Thus, the supermanifolds of the triplectic \cite{3pl}
and modified triplectic \cite{mod3pl} quantization schemes consist of the
variables $x^{i}=(\phi^{A},\bar{\phi}_{A})$ and $\theta_{a}^{i}=(\pi_{a}%
^{A},\phi_{Aa}^{\ast})$, where $\phi^{A}$ are the fields of the configuration
space of a general gauge theory; the antifields $\bar{\phi}_{A} $ are the
sources of the combined BRST--antiBRST symmetry; the antifields $\phi
_{Aa}^{\ast}$ are the sources of the BRST and antiBRST transformations; while
$\pi^{Aa}$ are auxiliary (gauge-fixing) fields. A superfield description
\cite{L} of extended BRST symmetry requires an extension of triplectic
supermanifolds \cite{3pl,mod3pl} by the additional (external) variables
$y^{i}=(\lambda^{A},J_{A})$ arising in the original $Sp(2)$-covariant scheme
\cite{BLT}, where $\lambda^{A}$ are auxiliary (gauge-fixing) fields, and
$J_{A}$ are the sources to the fields $\phi^{A}$. The realization of extended
BRST\ symmetry in general coordinates \cite{gl} is based on a tensor analysis
on supermanifolds with coordinates\emph{\ }$(x^{i},\theta_{a}^{i})$\emph{. }In
this section, we propose a superfield formulation of the analysis \cite{gl}.

\subsection{Superfields, Component Transformations}

Let us consider a superspace spanned by space-time coordinates and an
$Sp(2)$-doublet of anticommuting coordinates $\eta^{a}$. Any function
$f(\eta)$ has a component representation,
\[
f(\eta)=f_{0}+\eta^{a}f_{a}+\eta^{2}f_{3},\;\;\;\eta^{2}\equiv\frac{1}{2}%
\eta_{a}\eta^{a},
\]
and an integral representation,
\[
f(\eta)=\int d^{2}\,\eta^{\prime}\,\delta(\eta^{\prime}-\eta)f(\eta^{\prime
}),\;\;\;\delta(\eta^{\prime}-\eta)=(\eta^{\prime}-\eta)^{2},
\]
where integration over $\eta^{a}$ is given by
\[
\int d^{2}\eta=0,\;\;\;\int d^{2}\eta\;\eta^{a}=0,\;\;\;\int d^{2}\eta
\;\eta^{a}\eta^{b}=\varepsilon^{ab}.
\]
In particular, for any superfield $f(\eta)$ we have
\[
\int d^{2}\eta\;\frac{\partial f(\eta)}{\partial\eta^{a}}=0,
\]
which implies the property of integration by parts
\[
\int d^{2}\eta\;\frac{\partial f(\eta)}{\partial\eta^{a}}g(\eta)=-\int
d^{2}\eta\;(-1)^{\varepsilon(f)}f(\eta)\frac{\partial g(\eta)}{\partial
\eta^{a}}\,,
\]
where derivatives with respect to $\eta^{a}$ are taken from the left.

Let us now introduce a set of superfields $z^{i}{(\eta)}$, $\epsilon
(z^{i})=\epsilon_{i}$, $i=1,...,N$, with the component notation
\[
z^{i}{(\eta)=x}^{i}+\eta^{a}\theta_{a}^{i}+\eta^{2}y^{i},
\]
and the following distribution of Grassmann parity:
\[
\epsilon({x}^{i})=\epsilon(y^{i})=\epsilon_{i},\;\;\;\epsilon(\theta_{a}%
^{i})=\epsilon_{i}+1.
\]

We shall identify the components $(x^{i},\theta_{a}^{i},y^{i})$ with local
coordinates of a supermanifold $\mathcal{N}$, $\mathrm{dim\,}\,\mathcal{N}=4N
$, where the submanifold $\mathcal{M}$, $\mathrm{dim\,}\,\mathcal{M}=3N$, with
coordinates $(x^{i},\theta_{a}^{i})$ is chosen as a \textit{triplectic
supermanifold} \cite{gln,gl}. We accordingly define the following
transformations of the local coordinates:%

\begin{equation}
\bar{x}^{i}=\bar{x}^{i}(x),\;\;\;\bar{\theta}_{a}^{i}=\theta_{a}^{j}%
\frac{\partial\bar{x}^{i}}{\partial{x}^{j}}\,,\;\;\;\bar{y}^{i}=y^{i}\,,
\label{tr_trans}%
\end{equation}
where $\bar{x}^{i}=\bar{x}^{i}(x)$ are transformations on the submanifold $M$,
$\mathrm{dim}\,M=N$, with coordinates $(x^{i})$, called the \textit{base
supermanifold} \cite{gl}. The transformations of the coordinates
$(x^{i},\theta_{a}^{i})$ are identical with the transformations which define a
triplectic supermanifold \cite{gln,gl}. The superfield derivative
$\frac{\overleftarrow{\partial}}{\partial z^{i}{(\eta)}}$ with respect to
variations $\delta z^{i}(\eta)=\delta{x}^{i}+\eta^{a}\delta\theta_{a}^{i}$
induced by the component transformations (\ref{tr_trans}),
\begin{equation}
\frac{\overleftarrow{\partial}}{\partial z^{i}{(\eta)}}=\frac{\overleftarrow
{\partial}}{\partial\theta_{a}^{i}}\eta_{a}+\frac{\overleftarrow{\partial}%
}{\partial x^{i}}\eta^{2}\,, \label{sup_deriv}%
\end{equation}
is trivial on the external variables $y^{i}$. Using the derivative
(\ref{sup_deriv}) and the transformations (\ref{tr_trans}), we can introduce a
superfield extension of covariant differentiation \cite{gln,gl} on triplectic
supermanifolds $\mathcal{M}$.

\subsection{Superfield Extension of Triplectic Covariant Derivative}

As a preliminary step, we shall discuss some elements of tensor analysis on
the base supermanifold $M$, referring for a detailed treatment of
supermanifolds to the monograph \cite{DeWitt}. To this end, let us consider a
local coordinate system $(x)=(x^{1},...,x^{N})$ on the base supermanifold $M$,
in the vicinity of a point $P$. Let the sets $\{e_{i}\}$ and $\{e^{i}\}$ be
coordinate bases in the tangent space $T_{P}M$ and the cotangent space
$T_{P}^{\ast}M$, respectively. Under a change of coordinates $(x)\rightarrow
({\bar{x}})$, the basis vectors in $T_{P}M$ and $T_{P}^{\ast}M$ transform
according to
\[
{\bar{e}}^{i}=e^{j}\frac{\partial{\bar{x}}^{i}}{\partial x^{j}}\,,\;\;\;{\bar
{e}}_{i}=e_{j}\frac{\partial_{r}x^{j}}{\partial{\bar{x}}^{i}}\,.
\]
The transformation matrices obey the following relations:
\[
\frac{\partial_{r}{\bar{x}}^{i}}{\partial x^{k}}\frac{\partial_{r}x^{k}%
}{\partial{\bar{x}}^{j}}=\delta_{j}^{i}\,,\;\;\;\frac{\partial x^{k}}%
{\partial{\bar{x}}^{j}}\frac{\partial{\bar{x}}^{i}}{\partial x^{k}}=\delta
_{j}^{i}\,,\;\;\;\frac{\partial_{r}x^{i}}{\partial{\bar{x}}^{k}}\frac
{\partial_{r}{\bar{x}}^{k}}{\partial x^{j}}=\delta_{j}^{i}\,,\;\;\;\frac
{\partial{\bar{x}}^{k}}{\partial x^{j}}\frac{\partial x^{i}}{\partial{\bar{x}%
}^{k}}=\delta_{j}^{i}\,.
\]

A tensor field of type $(n,m)$ with rank $n+m$ is given by a set of functions
$T_{\;\;\;\;\;\;\;\;\ j_{1}...j_{m}}^{i_{1}\ldots i_{n}}(x)$, with Grassmann
parity $\epsilon(T_{\;\;\;\;\;\;\;\;\ j_{1}...j_{m}}^{i_{1}\ldots i_{n}%
})=\epsilon(T)+\epsilon_{i_{1}}+\cdot\cdot\cdot+\epsilon_{i_{n}}%
+\epsilon_{j_{1}}+\cdot\cdot\cdot+\epsilon_{j_{m}}\,$, which transform under a
change of coordinates, $(x)\rightarrow({\bar{x}})$, according to
\begin{align}
\bar{T}_{\;\;\;\;\;\;\;\;\ j_{1}...j_{m}}^{i_{1}...i_{n}}  &
=T_{\;\;\;\;\;\;\;\;\ k_{1}...k_{m}}^{l_{1}...l_{n}}\frac{\partial_{r}%
x^{k_{m}}}{\partial{\bar{x}}^{j_{m}}}\cdot\cdot\cdot\frac{\partial_{r}%
x^{k_{1}}}{\partial{\bar{x}}^{j_{1}}}\frac{\partial{\bar{x}}^{i_{n}}}{\partial
x^{l_{n}}}\cdot\cdot\cdot\frac{\partial{\bar{x}}^{i_{1}}}{\partial x^{l_{1}}%
}\nonumber\\
&  \times(-1)^{\left(  \sum\limits_{s=1}^{m-1}\sum\limits_{p=s+1}^{m}%
\epsilon_{j_{p}}(\epsilon_{j_{s}}+\epsilon_{k_{s}})+\sum\limits_{s=1}^{n}%
\sum\limits_{p=1}^{m}\epsilon_{j_{p}}(\epsilon_{i_{s}}+\epsilon_{l_{s}}%
)+\sum\limits_{s=1}^{n-1}\sum\limits_{p=s+1}^{n}\epsilon_{i_{p}}%
(\epsilon_{i_{s}}+\epsilon_{l_{s}})\right)  }. \label{tenzor}%
\end{align}
In particular, it is easy to see that the unit matrix $\delta_{j}^{i}$\ is a
tensor field of type $(1,1)$.

By analogy with tensor analysis on manifolds, on supermanifolds one introduces
an operation $\nabla\equiv\overleftarrow{\nabla}$ of covariant differentiation
of tensor fields, by the requirement that this operation should map a tensor
field of type $(n,m)$ into a tensor field of type $(n,m+1)$, and that, in case
one can introduce local Cartesian coordinates, it should reduce to the usual
differentiation. On an arbitrary supermanifold $M$, a covariant derivative is
given by a variety of differentiations with respect to separate coordinates,
$\nabla=(\overset{M}{\nabla}_{i})$. These differentiations are local
operations, acting on a tensor field of type $(n,m)$ by the rule
\begin{align}
T_{\;\;\;\;\;\;\;\;\ j_{1}...j_{m}}^{i_{1}\ldots i_{n}}\overset{M}{\nabla
}_{k}  &  =T_{\;\;\;\;\;\;\;\;\ j_{1}...j_{m},k}^{i_{1}\ldots i_{n}}%
+\underset{r=1}{\overset{n}{\sum}}\,T_{\;\;\;\;\;\;\;\;\;\;\;\;\;\;j_{1}%
...j_{m}}^{i_{1}...l...j_{n}}\,\overset{M}{\Gamma}\,_{\;\;\;lk}^{i_{r}%
}(-1)^{(\epsilon_{i_{r}}+\epsilon_{l})\left(  \epsilon_{l}+\overset
{n}{\underset{p=r+1}{\sum}}\epsilon_{i_{p}}+\underset{p=1}{\overset{m}{\sum}%
}\epsilon_{j_{p}}\right)  }\nonumber\\
&  -\underset{s=1}{\overset{m}{\sum}\,}T_{\;\;\;\;\;\;\;\;\ j_{1}...l...j_{m}%
}^{i_{1}\ldots i_{n}}\,\overset{M}{\Gamma}\,_{\;\,j_{s}k}^{l}(-1)^{(\epsilon
_{j_{s}}+\epsilon_{l})\underset{p=s+1}{\overset{m}{\sum}}\epsilon_{j_{p}}},
\label{m_nabla}%
\end{align}
where $\overset{M}{\Gamma}\,_{\;\;\;ij}^{k}(x)$ are generalized Christoffel
symbols (connection coefficients), subject to the transformation law
\[
\overset{M}{\bar{\Gamma}}\,_{\;\;\;ij}^{k}=(-1)^{\epsilon_{j}(\epsilon
_{m}+\epsilon_{i})}\frac{\partial_{r}\bar{x}^{k}}{\partial x^{l}}\overset
{M}{\Gamma}\,_{\;\;mn}^{l}\frac{\partial_{r}x^{n}}{\partial\bar{x}^{j}}%
\frac{\partial_{r}x^{m}}{\partial\bar{x}^{i}}+\frac{\partial_{r}\bar{x}^{k}%
}{\partial x^{m}}\frac{\partial_{r}^{2}x^{m}}{\partial\bar{x}^{i}\partial
\bar{x}^{j}}\,.
\]
In this paper, we restrict the consideration to \textit{symmetric}
connections, i.e., those possessing the property
\[
\overset{M}{\Gamma}\,_{\;\;ij}^{k}=(-1)^{\epsilon_{i}\epsilon_{j}}\overset
{M}{\Gamma}\,_{\;\;ji}^{k}\,.
\]
Note that this property is fulfilled automatically in case a local Cartesian
system can be introduced on the supermanifold $M$.

The curvature tensor $\overset{M}{R}\,_{\;\;mjk}^{i}(x)$ is defined by the
action of the (generalized) commutator of covariant derivatives $[\overset
{M}{\nabla}_{i},\overset{M}{\nabla}_{j}]=\overset{M}{\nabla}_{i}\overset
{M}{\nabla}_{j}-(-1)^{\epsilon_{i}\epsilon_{j}}\overset{M}{\nabla}_{j}%
\overset{M}{\nabla}_{i}$ on a vector field $T^{i}$ by the rule
\[
T^{i}[\overset{M}{\nabla}_{j},\overset{M}{\nabla}_{k}]=-(-1)^{\epsilon
_{m}(\epsilon_{i}+1)}T^{m}\overset{M}{R}\,_{\;\;mjk}^{i}\,.
\]
A straightforward calculation yields the following result:
\begin{equation}
\overset{M}{R}\,_{\;\;mjk}^{i}=-\overset{M}{\Gamma}\,_{\;\;mj,k}^{i}%
+\overset{M}{\Gamma}\,_{\;\;mk,j}^{i}(-1)^{\epsilon_{j}\epsilon_{k}}%
+\overset{M}{\Gamma}\,_{\;\;jl}^{i}\overset{M}{\Gamma}\,_{\;\;mk}%
^{l}(-1)^{\epsilon_{j}\epsilon_{m}}-\overset{M}{\Gamma}\,_{\;\;kl}^{i}%
\overset{M}{\Gamma}\,_{\;\;mj}^{l}(-1)^{\epsilon_{k}(\epsilon_{m}+\epsilon
_{j})}. \label{R}%
\end{equation}
The curvature tensor (\ref{R}) possesses the property of generalized symmetry
\[
\overset{M}{R}\,_{\;\;mjk}^{i}=-(-1)^{\epsilon_{j}\epsilon_{k}}\overset{M}%
{R}\,_{\;\;mkj}^{i}%
\]
and obeys the Jacobi identity
\[
(-1)^{\epsilon_{j}\epsilon_{l}}\overset{M}{R}\,_{\;\;jkl}^{i}+\mathrm{cycle\,}%
(j,k,l)\equiv0.
\]

On triplectic supermanifolds $\mathcal{M}$, one defines \cite{gl} covariant
differentiation of tensor fields transforming as tensors on the base
supermanifold $M$. In a similar way, we introduce a superfield extension of
the triplectic covariant derivative. Having in mind the coordinate
transformations (\ref{tr_trans}) on the supermanifold $\mathcal{N}$, we define
a tensor field of type $(n,m)$ and rank $n+m$ as a geometric object which in
any local coordinate system $(x,\theta,y)$ is given by a set of functions
$T_{\;\;\;\;\;\;\;\;\ j_{1}\ldots j_{m}}^{i_{1}\ldots i_{n}}(z)$ transforming
by the tensor law (\ref{tenzor}). Let us define the superfield covariant
derivative $\mathcal{D}\equiv\overleftarrow{\mathcal{D}}$ in a Cartesian
coordinate system to coincide with $\frac{\overleftarrow{\partial}}{\partial
z^{i}(\eta)}$, given by (\ref{sup_deriv}). Then, in general coordinates,
$\mathcal{D}=(\mathcal{D}_{i}(\eta))$ becomes
\[
\overleftarrow{\mathcal{D}}_{i}(\eta)=\frac{\overleftarrow{\partial}}%
{\partial\theta_{a}^{i}}\,\eta_{a}+\overset{\mathcal{M}}{\overleftarrow
{\nabla}}_{i}\,\eta^{2}.
\]
Here, each term of the $\eta$-expansion acts as a covariant differentiation of
tensor fields $T_{\;\;\;\;\;\;\;\;\ j_{1}\ldots j_{m}}^{i_{1}\ldots i_{n}}%
(z)$. The component $\overset{\mathcal{M}}{\nabla}_{i}$\thinspace\thinspace is
an extension of the covariant derivative $\overset{M}{\nabla}_{i}$, given by
(\ref{m_nabla}) on the base supermanifold $M$, namely,
\begin{equation}
\overset{\mathcal{M}}{\overleftarrow{\nabla}}_{i}=\overset{M}{\overleftarrow
{\nabla}}_{i}-\frac{\overleftarrow{\partial}}{\partial\theta_{a}^{k}}%
\theta_{a}^{m}\overset{M}{\Gamma}\,_{\;\;mi}^{k}(-1)^{\epsilon_{m}%
(\epsilon_{k}+1)}. \label{tcov}%
\end{equation}
The operation $\overset{\mathcal{M}}{\nabla}_{i}$ coincides\footnote{To
observe the coincidence of (\ref{tcov}) with the triplectic covariant
derivative \cite{gl}, one should go over to the parameterization
$(x^{i},\theta_{ia})$, where $\theta_{ia}$ transform as vectors of the tangent
space $T_{P}M$ (for details, see Section 3.4).} with the triplectic covariant
derivative \cite{gl}.

Since by definition $(x^{i},\theta_{a}^{i})$ are independent coordinates,
(\ref{m_nabla}), (\ref{tcov}) imply that the vectors $\theta_{a}^{i}$ are
covariantly constant with respect to $\overset{\mathcal{M}}{\nabla}_{i}$,
namely,
\begin{equation}
\theta_{a}^{i}\overset{\mathcal{M}}{\nabla}_{j}=0. \label{thecov}%
\end{equation}
By virtue of (\ref{m_nabla}), (\ref{tcov}), the commutator of two superfield
covariant derivatives $\mathcal{D}_{i}(\eta)$ has the form
\[
\left[  \mathcal{D}_{i}(\eta),\mathcal{D}_{j}(\eta^{\prime})\right]
=[\overset{\mathcal{M}}{\nabla}_{i},\overset{\mathcal{M}}{\nabla}_{j}%
]\,\eta^{2}\left(  \eta^{\prime}\right)  ^{2}.
\]
>From (\ref{tcov}), (\ref{thecov}), it follows that the action of this
commutator on a scalar field $T=T(z)$ is given by
\[
T\left[  \mathcal{D}_{i}(\eta),\mathcal{D}_{j}(\eta^{\prime})\right]
=(-1)^{\epsilon_{m}(\epsilon_{n}+1)}\eta^{2}\left(  \eta^{\prime}\right)
^{2}\frac{\partial_{r}T}{\partial\theta_{a}^{n}}\theta_{a}^{m}\overset{M}%
{R}\,_{\;\;mij}^{n}\,,
\]
where $\overset{M}{R}\,\,_{\;\;mij}^{n}$ is the curvature tensor (\ref{R}) on
the base supermanifold.

\section{Superfield Realization of (Modified) Triplectic Algebra}

The extended BRST\ quantization in general coordinates \cite{gl} is based on a
realization of the so-called triplectic \cite{3pl} and modified triplectic
\cite{mod3pl} operator algebras. The operators obeying these algebras are
originally defined on triplectic supermanifolds $\mathcal{M}$. In this
section, we propose a superfield formulation of \cite{gl}, realized on
extended supermanifolds $\mathcal{N}$. Namely, we construct a manifestly
superfield realization of the (modified) triplectic algebra, which permits us
to formulate a superfield realization of extended BRST\ quantization in
general coordinates, along the lines of \cite{gl}.

\subsection{Triplectic and Modified Triplectic Algebras}

The triplectic algebra \cite{3pl} includes two sets of second- and first-order
operators, $\overleftarrow{\Delta}^{a}$ and $\overleftarrow{V}^{a}$,
respectively, having the Grassmann parity $\epsilon(\Delta^{a})=\epsilon
(V^{a})=1$, and obeying the following relations:%

\begin{equation}
\Delta^{\{a}\Delta^{b\}}=0,\;\;\;V^{\{a}V^{b\}}=0,\;\;\;V^{a}\Delta^{b}%
+\Delta^{b}V^{a}=0. \label{dd}%
\end{equation}
The modified triplectic quantization \cite{mod3pl}, in comparison with the
$Sp(2)$-covariant approach \cite{BLT} and the triplectic scheme \cite{3pl},
involves an additional $Sp(2)$-doublet of first-order operators
$\overleftarrow{U}^{a}$, $\epsilon(U^{a})=1$, with the modified triplectic
algebra \cite{mod3pl} given by the relations
\begin{gather}
\Delta^{\{a}\Delta^{b\}}=0,\;\;\;V^{\{a}V^{b\}}=0,\;\;\;U^{\{a}U^{b\}}%
=0,\nonumber\\
V^{\{a}\Delta^{b\}}+\Delta^{\{b}V^{a\}}=0,\;\;\;\Delta^{\{a}U^{b\}}%
+U^{\{a}\Delta^{b\}}=0,\;\;\;U^{\{a}V^{b\}}+V^{\{a}U^{b\}}=0. \label{mal}%
\end{gather}
In (\ref{dd}), (\ref{mal}), the curly brackets denote symmetrization with
respect to the enclosed indices $a$ and $b$.

Using the odd second-order differential operators $\Delta^{a}$, one can
introduce a pair of bilinear operations $(\;\,,\;)^{a}$, by the rule
\begin{equation}
(F,G)^{a}=(-1)^{\epsilon(G)}(FG)\Delta^{a}-(-1)^{\epsilon(G)}(F\Delta
^{a})G-F(G\Delta^{a}). \label{ab}%
\end{equation}
The operations (\ref{ab}) possess the Grassmann parity $\epsilon
((F,G)^{a})=\epsilon(F)+\epsilon(G)+1$ and obey the following symmetry
property:
\[
(F,G)^{a}=-(-1)^{(\epsilon(G)+1)(\epsilon(F)+1)}(G,F)^{a}.
\]
The operations (\ref{ab}) are linear with respect to both arguments,
\[
(F+G,H)^{a}=(F,H)^{a}+(G,H)^{a},\quad(F,G+H)^{a}=(F,G)^{a}+(F,H)^{a},
\]
and obey the Leibniz rule
\[
(F,GH)^{a}=(F,G)^{a}H+(F,H)^{a}G(-1)^{\epsilon(G)\epsilon(H)}.
\]
Due to the properties (\ref{dd}) of the operators $\Delta^{a}$, the odd
bracket operations satisfy the generalized Jacobi identity
\[
(F,(G,H)^{\{a})^{b\}}(-1)^{(\epsilon(F)+1)(\epsilon(H)+1)}+\mathrm{cycle}%
(F,G,H)\equiv0.
\]

In view of their properties, the operations $(\;,\;)^{a}$ form a set of
antibrackets, such as those introduced for the first time in \cite{BLT}.
Therefore, having an explicit realization of operators $\Delta^{a}$ with the
properties (\ref{dd}), one can generate the extended antibrackets explicitly,
using (\ref{ab}). Explicit realizations of $\Delta^{a}$ are known in two
cases: in Darboux coordinates \cite{BLT,3pl,mod3pl}, and in general
coordinates on triplectic supermanifolds $\mathcal{M}$, where the base
supermanifold $M$ is a flat Fedosov supermanifold \cite{gl,gl2}.

\subsection{Realization of Triplectic Algebra}

To find an explicit superfield realization of the triplectic algebra
(\ref{dd}) in general coordinates, we shall use the assumptions of \cite{gl}
concerning the properties of the base supermanifold $M$. Thus, we equip $M$
with a Poisson structure, namely, with a nondegenerate \emph{even} second-rank
tensor field $\omega^{ij}(x)$, and its inverse $\omega_{ij}(x)$,
$\epsilon(\omega^{ij})=\epsilon(\omega_{ij})=\epsilon_{i}+\epsilon_{j}$,
\[
\omega^{ik}\omega_{kj}(-1)^{\epsilon_{k}}=\delta_{j}^{i}\,,\;\;\;\omega
_{ik}\omega^{kj}(-1)^{\epsilon_{i}}=\delta_{i}^{j}\,,
\]
obeying the properties of generalized antisymmetry
\[
\omega^{ij}=-(-1)^{\epsilon_{i}\epsilon_{j}}\omega^{ji}\Leftrightarrow
\omega_{ij}=-(-1)^{\epsilon_{i}\epsilon_{j}}\omega_{ji},
\]
and satisfying the following Jacobi identities:
\[
\omega^{il}\partial_{l}\omega^{jk}(-1)^{\epsilon_{i}\epsilon_{k}%
}+\mathrm{cycle}(i,j,k)\equiv0\Leftrightarrow\omega_{ij,k}(-1)^{\epsilon
_{i}\epsilon_{k}}+\mathrm{cycle}(i,j,k)\equiv0.
\]
The tensor field $\omega^{ij}$ defines a Poisson bracket \cite{gl}, and, due
to its nondegeneracy, also a corresponding \emph{even }symplectic structure
\cite{gl} on the base supermanifold. In view of this fact, the supermanifold
$M$ can be regarded as an even Poisson supermanifold, as well as an even
symplectic supermanifold. Following \cite{gl}, we demand that the covariant
derivative $\overset{M}{\nabla}_{i}$ should respect the Poisson structure
$\omega^{ij}$,
\begin{equation}
\omega^{ij}\overset{M}{\nabla}_{k}=0\Leftrightarrow\omega_{ij}\overset
{M}{\nabla}_{k}=0, \label{covom}%
\end{equation}
which provides the covariant constancy of the differential two-form
$\omega=\omega_{ij}dx^{j}\wedge dx^{i}$. Thus, the base supermanifold $M$ can
be regarded as an even symplectic supermanifold, being a supersymmetric
extension \cite{gl,gl2} of the Fedosov manifold \cite{F,fm}. One can formally
identify $\omega^{ij}$ and $\omega_{ij}$ with some functions of the
supervariables $z^{i}(\eta)$, i.e., $\Omega^{ij}(z)=\omega^{ij}(x)$ and
$\Omega_{ij}(z)=\omega_{ij}(x)$. It is obvious that the tensor fields
$\Omega^{ij}$ and $\Omega_{ij}$ are covariantly constant:
\[
\Omega^{ij}\mathcal{D}_{k}(\eta)=\Omega_{ij}\mathcal{D}_{k}(\eta)=0.
\]

The introduced structures allow one to equip the supermanifold $\mathcal{N}$
with a superfield $Sp(2)$-irreducible second-rank tensor $S_{ab}$,
\begin{equation}
S_{ab}=\frac{1}{6}\int d^{2}\eta\,{\eta}^{2}\frac{\partial z^{i}}{\partial
\eta^{a}}\Omega_{ij}\frac{\partial z^{j}}{\partial\eta^{b}}\,,\;\;\;\epsilon
(S_{ab})=0, \label{Sab}%
\end{equation}
invariant under changes of local coordinates on $\mathcal{N}$, i.e., ${\bar
{S}}_{ab}=S_{ab}$, and symmetric with respect to the $Sp(2)$-indices,
$S_{ab}=S_{ba}$.

Following \cite{gln,gl}, we also equip the base supermanifold $M$ with a
scalar density $\rho(x)$, $\epsilon(\rho)=0$. Using the covariant derivative
$\mathcal{D}_{i}(\eta)$, we can construct a superfield $Sp(2)$-doublet of odd
second-order differential operators $\Delta^{a}$, acting as scalars on the
supermanifold $\mathcal{N}$,
\begin{equation}
\overleftarrow{\Delta}^{a}=\int d^{2}\eta\,{\eta}^{2}\left(  \overleftarrow
{\mathcal{D}}_{i}\frac{\partial_{r}}{\partial\eta^{a}}\right)  \Omega
^{ij}\left[  \left(  \overleftarrow{\mathcal{D}}_{j}+\frac{1}{2}%
(\mathcal{R}\overleftarrow{\mathcal{D}}_{j})\right)  \frac{\partial_{r}%
}{\partial{\eta}^{2}}\right]  (-1)^{\epsilon_{i}+\epsilon_{j}}, \label{Delt}%
\end{equation}
where $\mathcal{R}(z)\equiv\rho(x)$.

The operators (\ref{Delt}) generate a superfield $Sp(2)$-doublet of
antibracket operations,
\begin{equation}
(F,G)^{a}=-\int d^{2}\eta\,\eta^{2}\left(  F\mathcal{D}_{i}\frac{\partial_{r}%
}{\partial{\eta}^{2}}\right)  \Omega^{ij}\frac{\partial}{\partial\eta^{a}%
}\left(  G\mathcal{D}_{j}\right)  (-1)^{\epsilon_{j}\epsilon(G)}%
-(-1)^{(\epsilon(F)+1)(\epsilon(G)+1)}(F\leftrightarrow G). \label{2op}%
\end{equation}
These operations possess all the properties of extended antibrackets
\cite{BLT}, except the Jacobi identity, which is closely related to the
properties (\ref{dd}) of anticommutativity and nilpotency of $\Delta^{a}$.

Using the operations (\ref{2op}) and the irreducible second-rank
$Sp(2)$-tensor $S_{ab}$ in (\ref{Sab}), we define the following $Sp(2)$%
-doublet of odd first-order differential operators $V_{a}$:
\begin{equation}
\overleftarrow{V}_{a}=(\;\cdot,S_{ab})^{b}=-\frac{1}{2}\int d^{2}\eta
\,\eta^{2}\left(  \overleftarrow{\mathcal{D}}_{i}\frac{\partial_{r}}%
{\partial\eta^{2}}\right)  \frac{\partial_{r}z^{i}}{\partial\eta^{a}}\,.
\label{Va}%
\end{equation}

Straightforward calculations, analogous to \cite{gl}, with allowance for the
manifest form of the operators $\Delta^{a}$, $V^{a}$, (\ref{Delt}),
(\ref{Va}), show that there exists such a choice of the density function
$\mathcal{R}$,
\[
\mathcal{R}=-\mathrm{log}\;\mathrm{sdet}\;\left(  \Omega^{ij}\right)  ,
\]
that the triplectic algebra (\ref{dd}) is fulfilled on $\mathcal{N}$ in case
the base supermanifold $M$ is a flat Fedosov supermanifold:
\[
\overset{M}{R}\,_{\;\;mjk}^{i}=0,
\]
with the curvature tensor $\overset{M}{R}\,_{\;\;mjk}^{i}$ given by (\ref{R}).
Thus, we have explicitly realized the extended antibrackets (\ref{2op}) and
the triplectic algebra (\ref{dd}) of the generating operators $\Delta^{a} $,
$V^{a}$.

\subsection{Realization of Modified Triplectic Algebra}

In view of (\ref{dd}), to complete the explicit superfield realization of the
modified triplectic algebra (\ref{mal}) in general coordinates, it remains to
construct the operators $U^{a}$. To this end, following \cite{gl}, we
introduce another geometrical structure on the base supermanifold $M$. Namely,
we consider a symmetric second-rank tensor $g_{ij}(x)=(-1)^{\epsilon
_{i}\epsilon_{j}}g_{ji}(x)$, which we identify with a tensor field $G_{ij}%
(z)$. The introduced tensor field can be used to construct on $\mathcal{N}$ an
$Sp(2)$ scalar function $S_{0}$, the so-called anti-Hamiltonian,
\begin{equation}
S_{0}=\frac{1}{2}\varepsilon^{ab}\int d^{2}\eta\,\eta^{2}\frac{\partial
_{r}z^{i}}{\partial\eta^{a}}\,G_{ij}\frac{\partial_{r}z^{j}}{\partial\eta^{b}%
}\,,\;\;\;\epsilon(S_{0})=0. \label{S0}%
\end{equation}
The anti-Hamiltonian $S_{0}$ generates vector fields $U^{a}$,
\begin{align*}
\overleftarrow{U}^{a}  &  =(\;\cdot,S_{0})^{a}=\int d^{2}\eta\,\eta^{2}\left[
\left(  \overleftarrow{\mathcal{D}}_{i}\frac{\partial_{r}}{\partial\eta^{2}%
}\right)  \Omega^{im}G_{mn}\frac{\partial z^{n}}{\partial\eta_{a}}\right.
(-1)^{\epsilon_{m}}\\
&  \;\;\;\;\;\;\;\;\;\;\;\;\;\;\;\;\;+\left.  \frac{1}{2}\left(
\overleftarrow{\mathcal{D}}_{i}\frac{\partial_{r}}{\partial\eta^{a}}\right)
\Omega^{ij}\frac{\partial_{r}z^{m}}{\partial\eta^{c}}\left(  G_{mn}%
\overleftarrow{\mathcal{D}}_{j}\frac{\partial_{r}}{\partial\eta^{2}}\right)
\frac{\partial_{r}z^{n}}{\partial\eta_{c}}(-1)^{\epsilon_{i}+\epsilon
_{j}\epsilon_{n}}\right]  .
\end{align*}
The algebraic conditions (\ref{mal}) yield the following equations for $S_{0}%
$:
\begin{equation}
(S_{0},S_{0})^{a}=0,\;\;\;S_{0}V^{a}=0,\;\;\;S_{0}\Delta^{a}=0. \label{S1}%
\end{equation}
Solutions of these equations always exist. An example of such soultions can be
found in the class of covariantly constant\footnote{In the class of
covariantly constant tensors $G_{ij}$, solutions of (\ref{S1}) can be selected
by imposing the condition $G_{ij}\left(  \mathcal{RD}_{k}\right)  \Omega
^{kj}=0$. The simplest solution of this kind is given by a covariantly
constant scalar density $\mathcal{R}.$} tensor fields $G_{ij}$, $G_{ij}%
\mathcal{D}_{k}=0$. We do not restrict ourselves to this special case, and
simply assume that equations (\ref{S1}) are fulfilled. Thus, we obtain a
realization of the modified triplectic algebra (\ref{mal}), and have at our
disposal all the ingredients for the quantization of general gauge theories
within the modified triplectic scheme.

\subsection{Quantization}

The quantization procedure repeats all the essential steps taken for the first
time in \cite{gln}, and leads to the vacuum functional
\begin{equation}
Z=\int dz\,\mathcal{D}_{0}\,\mathrm{exp}{\{(i/\hbar)[W+X+\alpha S_{0}]\}},
\label{Z}%
\end{equation}
where $\alpha$ is an arbitrary constant; the function $S_{0}$ is given by
(\ref{S0}), while the quantum action $W=W(z)$ and the gauge-fixing functional
$X=X(z)$ satisfy the following quantum master equations:
\begin{align}
\frac{1}{2}(W,W)^{a}+W\mathcal{V}^{a}  &  =i\hbar W\Delta^{a},\label{MEW}\\
\frac{1}{2}(X,X)^{a}+X\mathcal{U}^{a}  &  =i\hbar X\Delta^{a}. \label{MEX}%
\end{align}
In (\ref{Z}), integration over the supervariables is understood as integration
over their components,
\[
dz=dx\,d\theta_{a}\,dy,
\]
with the integration measure $\mathcal{D}_{0}$ given by
\[
\mathcal{D}_{0}=\left[  \mathrm{sdet}\left(  \Omega^{ij}\right)  \right]
^{-3/2}.
\]
In (\ref{MEW}) and (\ref{MEX}), we have introduced operators $\mathcal{V}^{a}
$, $\mathcal{U}^{a}$, according to
\[
\mathcal{V}^{a}=\frac{1}{2}\left(  \alpha U^{a}+\beta V^{a}+\gamma
U^{a}\right)  ,\;\;\;\mathcal{U}^{a}=\frac{1}{2}\left(  \alpha U^{a}-\beta
V^{a}-\gamma U^{a}\right)  .
\]
It is obvious that for arbitrary constants $\alpha$, $\beta$, $\gamma$ the
operators $\mathcal{V}^{a}$, $\mathcal{U}^{a}$ obey the properties
\[
\mathcal{V}^{\{a}\mathcal{V}^{b\}}=0,\;\;\;\mathcal{U}^{\{a}\mathcal{U}%
^{b\}}=0,\;\;\;\mathcal{V}^{\{a}\mathcal{U}^{b\}}+\mathcal{U}^{\{a}%
\mathcal{V}^{b\}}=0.
\]
Therefore, the operators $\Delta^{a}$, $\mathcal{V}^{a}$, $\mathcal{U}^{a}$
also realize the modified triplectic algebra.

The integrand of the vacuum functional (\ref{Z}) is invariant under extended
BRST transformations defined by the generators
\begin{equation}
\delta^{a}=(\;\cdot,W-X)^{a}+\mathcal{V}^{a}-\mathcal{U}^{a}. \label{BRST}%
\end{equation}
In the usual manner, this allows one to prove that, for every given set of the
parameters $\alpha$, $\beta$, $\gamma$, the vacuum functional (\ref{Z}) does
not depend on a choice of the gauge-fixing function $X$.

Let us analyze the component structure of the proposed quantization scheme in
order to establish its relation with the modified triplectic quantization in
general coordinates \cite{gl}. To this end, note that the integration measure
$\mathcal{D}_{0}$ and the function $S_{0}$,
\[
S_{0}=\frac{1}{2}\varepsilon^{ab}\theta_{a}^{i}\,g_{ij}\theta_{b}%
^{j}(-1)^{\epsilon_{i}+\epsilon_{j}},
\]
coincide with the corresponding objects of \cite{gl}. The operators
$\Delta^{a}$, $V^{a}$, $U^{a}$ and antibrackets $(\;,\;)^{a}$ have the form
\begin{align*}
\overleftarrow{\Delta}^{a}  &  =(-1)^{\epsilon_{i}}\frac{\overleftarrow
{\partial}}{\partial\theta_{ia}}\left(  \overset{\mathcal{M}}{\overleftarrow
{\nabla}}_{i}+\frac{1}{2}\rho_{,i}\right)  ,\\
\overleftarrow{V}^{a}  &  =\frac{1}{2}\epsilon^{ab}\overset{\mathcal{M}%
}{\overleftarrow{\nabla}}_{i}\omega^{ij}\theta_{jb}\,,\\
\overleftarrow{U}^{a}  &  =-\overset{\mathcal{M}}{\overleftarrow{\nabla}}%
_{i}\omega^{im}g_{mn}\theta^{na}(-1)^{\epsilon_{m}}-\frac{1}{2}\frac
{\overleftarrow{\partial}}{\partial\theta_{ia}}\theta_{c}^{m}(g_{mn}%
\overset{\mathcal{M}}{\overleftarrow{\nabla}}_{i})\theta^{nc}(-1)^{\epsilon
_{n}(\epsilon_{i}+1)+\epsilon_{m}},\\
(F,G)^{a}  &  =(F\overset{\mathcal{M}}{\overleftarrow{\nabla}}_{i}%
)\frac{\partial G}{\partial\theta_{ia}}-(-1)^{(\epsilon(F)+1)(\epsilon
(G)+1)}(G\overset{\mathcal{M}}{\overleftarrow{\nabla}}_{i})\frac{\partial
F}{\partial\theta_{ia}}\,,
\end{align*}
where $\theta_{ia}$, defined by $\theta_{a}^{i}=\omega^{ij}\theta
_{ja}(-1)^{\epsilon_{i}}$, are covariantly constant covectors, $\theta
_{ia}\overset{\mathcal{M}}{\nabla}_{j}=0$, while $\frac{\overleftarrow
{\partial}}{\partial\theta_{ia}}$ transform as vectors. The above component
expressions imply that the operators $\Delta^{a}$, $V^{a}$, $U^{a}$ and
antibrackets coincide with the corresponding objects of \cite{gl}, which
follows from the coincidence of $\overset{\mathcal{M}}{\nabla}_{i}$ with the
triplectic covariant derivative \cite{gl}, given by
\[
\overset{\mathcal{M}}{\overleftarrow{\nabla}}_{i}=\overset{M}{\overleftarrow
{\nabla}}_{i}+\frac{\overleftarrow{\partial}}{\partial\theta_{ma}}\theta
_{ka}\overset{M}{\Gamma}\,_{\;\;mi}^{k}.
\]
In case local Cartesian coordinates can be introduced on $M$, the coincidence
of derivatives is automatic, while in the case of arbitrary connection
coefficients, the coincidence takes place since $M$ is a Fedosov
supermanifold, namely, due to (\ref{covom}). Equations (\ref{MEW}),
(\ref{MEX}) formally coincide with the master equations of \cite{gl}, because
the external variables $y^{i}$ enter only as arguments of $W(z)$ and $X(z)$.
In Darboux coordinates $(\tilde{z}^{\mu},y^{i})$, $y^{i}=(\lambda^{A},J_{A})$,
one can choose solutions of (\ref{MEW}), (\ref{MEX}) as solutions of the
master equations \cite{gl}, namely, $W=W(\tilde{z})$, $X=X(\tilde{z},\lambda
)$. Since in the coordinates $(\tilde{z}^{\mu},y^{i})$ the tensor $\omega
^{ij}$ can be chosen \cite{gln} such that $\mathcal{D}_{0}=\mathrm{const}$,
the vacuum functional (\ref{Z}) reduces to
\[
Z=\int d\tilde{z}\,d\lambda\,\mathrm{exp}{\{(i/\hbar)[W(\tilde{z})+X(\tilde
{z},\lambda)+\alpha S_{0}(\tilde{z})]\},}%
\]
which is identical with the vacuum functional \cite{gl}, written in Darboux coordinates.

\section{Conclusion}

In this paper, we have proposed a superfield realization of extended BRST
symmetry in general coordinates, along the lines of our recent works
\cite{gln,gl} on modified triplectic quantization. We have found an explicit
superfield realization of the modified triplectic algebra of generating
operators $\Delta^{a}$, $V^{a}$, $U^{a}$ on an extended supermanifold
$\mathcal{N}$, obtained from the triplectic supermanifold $\mathcal{M}$ by
adding external supervariables, which, in Darboux coordinates, can be
interpreted as sources $J_{A}$ to the fields and as auxiliary gauge-fixing
variables $\lambda^{A}$. The present study applies the essential ingredients
of \cite{gln,gl}, and has the same general features. Thus, the base
supermanifold $M$ of fields and antifields is a flat Fedosov supermanifold
equipped with a symmetric structure. As in \cite{gln,gl}, the formalism is
characterized by free parameters, ($\alpha$, $\beta$, $\gamma$), whose
specific choice in Darboux coordinates reproduces all the known schemes of
covariant quantization based on extended BRST symmetry (for details, see
\cite{gln}). Every specific choice of the free parameters ($\alpha$, $\beta$,
$\gamma$) yields a gauge-independent vacuum functional and, therefore, a gauge
independent $S$-matrix (see \cite{t}).

\textbf{Acknowledgments:}~The authors are grateful to D.V. Vassilevich for
stimulating discussions. The work was supported by Deutsche
Forschungsgemeinschaft (DFG), grant GE 696/7-1. D.M.G. acknowledges the
support of the foundations FAPESP, CNPq and DAAD. The work of P.M.L. was
supported by the Russian Foundation for Basic Research (RFBR), 02-02-04002,
03-02-16193, and by the President grant 1252.2003.2 for Support of Leading
Scientific Schools. P.Yu.M. is grateful to FAPESP, grant 02/00423-4.

\end{document}